\documentclass[twocolumn,english,footinbib,bibnotes,superscriptaddress]{revtex4-2}
\usepackage{graphicx}
\usepackage[usenames,dvipsnames]{color}
\usepackage{bbm}
\usepackage{bm}
\usepackage{amsmath}
\usepackage{amssymb}
\usepackage{times}
\usepackage[colorlinks=true,linkcolor=Blue,citecolor=Blue,urlcolor=Blue]{hyperref}
\usepackage[version=3]{mhchem}

\graphicspath{{figures/compressed/}}

\begin{document}

\title{Functional renormalization group study of the Kitaev-$\Gamma$ model on the honeycomb lattice\\and emergent incommensurate magnetic correlations}
\author{Finn Lasse Buessen}
\affiliation{Department of Physics, University of Toronto, Toronto, Ontario M5S 1A7, Canada}
\author{Yong Baek Kim}
\affiliation{Department of Physics, University of Toronto, Toronto, Ontario M5S 1A7, Canada}
\date{\today}

\begin{abstract}
The theoretical inception of the Kitaev honeycomb model has had defining influence on the experimental hunt for quantum spin liquids, bringing unprecedented focus onto the synthesis of materials with bond-directional interactions. 
Numerous Kitaev materials, which are believed to harbor ground states parametrically close to the Kitaev spin liquid, have been investigated since. 
Yet, in these materials the Kitaev interaction often comes hand in hand with off-diagonal $\Gamma$ interactions -- with the competition of the two potentially giving rise to a magnetically ordered ground state. 
In an attempt to aid future material investigations, we study the phase diagram of the spin-1/2 Kitaev-$\Gamma$ model on the honeycomb lattice. 
Employing a pseudofermion functional renormalization group approach which directly operates in the thermodynamic limit and captures the joint effect of thermal and quantum fluctuations, we unveil the existence of extended parameter regimes with emergent incommensurate magnetic correlations at finite temperature. 
We supplement our results with additional calculations on a finite cylinder geometry to investigate the impact of periodic boundary conditions on the incommensurate order, thereby providing a perspective on previous numerical studies on finite systems. 
\end{abstract}

\maketitle


\section{Introduction}
\label{sec:introduction}
Quantum spin liquids are intriguing phases of matter which can manifest when strong quantum fluctuations inhibit the formation of conventional magnetic order~\cite{Balents2010,Witczak-Krempa2014}. 
They are accompanied by massive long-range entanglement which enables them to possess unusual properties; the physical constituents of the system may fractionalize and give rise to an emergent gauge structure as well as to the formation of collective parton degrees of freedom with new, effective quantum numbers~\cite{Savary2017,Broholm2020}. 
Among the many candidate models to host a quantum spin liquid ground state, the Kitaev honeycomb model stands out. 
It is not only exactly solvable at zero temperature~\cite{Kitaev2006}, but it is also amenable to quasi-exact Monte Carlo studies at finite temperature~\cite{Nasu2014,Nasu2015}, thus giving unprecedented systematic insight into the fractionalization process in frustrated quantum magnets. 

Following the theoretical inception of the Kitaev model and its generalization to three-dimensional lattices~\cite{OBrien2016,Eschmann2020}, a flurry of activity commenced in an effort to identify materials which could potentially realize the microscopic model. 
A number of so-called Kitaev materials with strong bond-directional interactions have been found since~\cite{Jackeli2009,Rau2016,Trebst2017,Takagi2019}. 
Prominent examples include \ce{Na2IrO3}~\cite{Singh2010,Chaloupka2010}, $\alpha$-\ce{Li2IrO3}~\cite{Singh2012}, and $\alpha$-\ce{RuCl3}~\cite{Plumb2014,Banerjee2016,Banerjee2017,Kasahara2018,Chern2020b} on the honeycomb lattice and $\beta$-\ce{Li2IrO3} on the three-dimensional hyper-honeycomb lattice~\cite{Takayama2015,Lee2014,Kimchi2014,Lee2015,Kim2015,Lee2016,Schaffer2016,Huang2018,Choi2019}. 
All these materials, however, exhibit magnetic ordering at temperatures below $T \approx 10\,\mathrm{K}$ ($T\approx 38\,\mathrm{K}$ for the 3D compound) which is driven by competing interactions that deviate from the perfect Kitaev model; such additional interactions often involve higher-symmetry Heisenberg interactions or lower-symmetry off-diagonal interactions, commonly referred to as $\Gamma$ interactions in the literature~\cite{Winter2016,Winter2017}. 
To this day, it is subject of ongoing research to further refine the available palette of Kitaev material candidates. 
The magnetic transition temperature can be suppressed, for example, by chemical substitution of inter-layer \ce{Li}-atoms as demonstrated in \ce{Ag3LiIr2O6} and \ce{H3LiIr2O6}~\cite{Bahrami2019,Kitagawa2018}, although for the latter material it is argued that chemical disorder may play a relevant role in the suppression of magnetic order~\cite{Li2018,Yadav2018}. 

In order to provide guidance for further material synthesis and refinement, it is immensely valuable to understand the underlying microscopic spin model and its associated phase diagram. 
As a minimal model for the microscopic description, the Kitaev-$\Gamma$ model has received ample attention. 
It comprises both, Kitaev and $\Gamma$ interactions, which are believed to be the two dominant terms in $\alpha$-\ce{RuCl3}~\cite{Winter2016,Winter2017,Sears2020}. 
Previous model calculations indicate that multiple competing ground states are in close vicinity to each other in the part of the parameter space which is relevant for $\alpha$-\ce{RuCl3}, i.e. ferromagnetic Kitaev interactions ($K<0$) in conjunction with antiferromagnetic $\Gamma$ interactions ($\Gamma>0$) -- thereby rendering the theoretical analysis of the model extremely challenging~\cite{Rau2014,Gohlke2018,Gohlke2020}. 
Furthermore, in some of the Kitaev materials, including $\alpha$-\ce{Li2IrO3} and $\beta$-\ce{Li2IrO3}, the formation of incommensurate counter-rotating magnetic spirals has been reported, making their numerical simulation on finite-size model clusters notoriously difficult~\cite{Williams2016,Biffin2014,Rousochatzakis2018,Ducatman2018}. 

Most importantly, however, the connection between model calculations and experiments remained incomplete, since previous studies have only focused on the zero-temperature properties of the model and experiments are necessarily performed at finite temperature. 
The role of thermal fluctuations and their potential driving of an entropic magnetic ordering transition (order by disorder) has not yet been explored. 

\begin{figure}
\includegraphics[width=\linewidth]{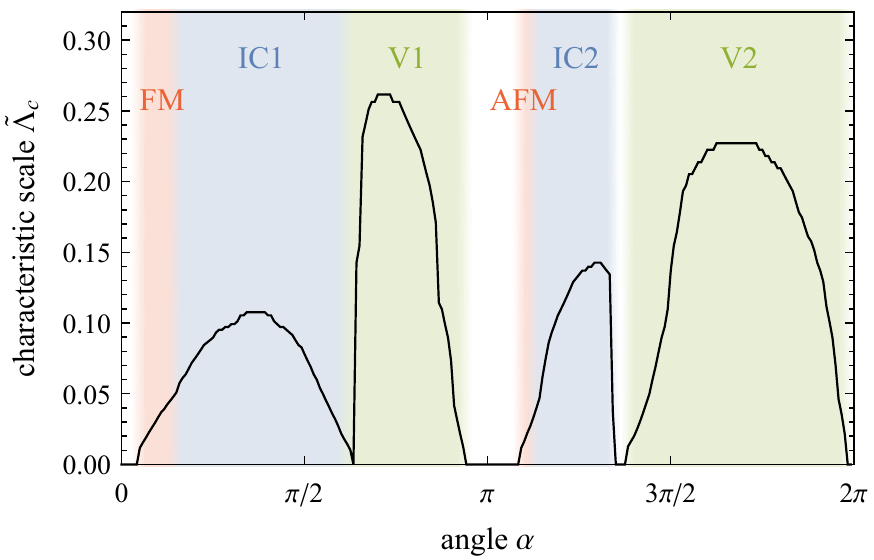}
\caption{{\bf Phase diagram} of the Kitaev-$\Gamma$ model on the honeycomb lattice as a function of the angle $\alpha$, which parametrizes the ratio of Kitaev coupling $K=-\cos(\alpha)$ and $\Gamma$ interaction $\Gamma=\sin(\alpha)$. The black curve indicates the characteristic RG scale $\widetilde{\Lambda}_{c}$ which is an indicator for the onset energy scale of the magnetic order, see text for details. The shaded regions indicate ordered phases with finite $\widetilde{\Lambda}_{c}$. Observed ordered phases are ferromagnetic order (FM), incommensurate phase 1 (IC1), vortex phase 1 (V1), antiferromagnetic order (AFM), incommensurate phase 2 (IC2), and vortex phase 2 (V2). White regions indicate spin liquid phases and the absence of spontaneous symmetry breaking, mainly the FM and AFM Kitaev spin liquids around $\alpha/\pi=0$ and $\alpha/\pi=1$, respectively.}
\label{fig:introduction:phasediagram}
\end{figure}

In this paper, we present a take on the spin-1/2 quantum Kitaev-$\Gamma$ model which is based on a pseudofermion functional renormalization group (pf-FRG) approach. 
The pf-FRG formalism operates on a genuinely infinite representation of the underlying honeycomb lattice and is thus able to faithfully resolve commensurate as well as incommensurate magnetic correlations, in contrast to approaches which operate on finite-size systems. 
Furthermore, the approach is sensitive to the impact of thermal and quantum fluctuations, thus complementing previous studies of the zero-temperature ground state. 
We study the phase diagram of the Kitaev-$\Gamma$ model and show that two extended phases exist which exhibit incommensurate spin correlations at finite temperature; in the context of the pf-FRG approach, these phases manifest as the first instability of the RG flow when the energy scale of the RG cutoff is lowered. 
For the incommensurate phases, we demonstrate that the dominant intensity peaks in the structure factor, which characterize the incommensurate order, shift continuously upon variation of the coupling constant ratio. 
Moreover, we identify magnetic vortex phases which are overshadowed by a sub-dominant incommensurability effect. 
The resulting phase diagram is summarized in Fig.~\ref{fig:introduction:phasediagram}.
In view of potential future studies, we point out that the pervasiveness of incommensurate correlations poses a challenge to many established methods which may rely on finite lattice representations; to substantiate this observation, we perform pf-FRG calculations on a finite cylinder geometry, assessing the extent to which an anisotropy is imposed by geometrical constraints. 
We find that this unphysical anisotropy can become sizable in phases with incommensurate magnetic correlations, whereas it remains negligible in the Kitaev spin liquid phases where correlations are short ranged. 

The structure of the paper is as follows. 
In Sec.~\ref{sec:model}, we define the Kitaev-$\Gamma$ model, set the notation, and review previous studies of the quantum model at zero temperature. 
We then briefly introduce the pf-FRG formalism in Sec.~\ref{sec:frg}, before we discuss our findings for the phase diagram in Sec.~\ref{sec:phasediagram}. 
We further focus on the incommensurate phases in Sec.~\ref{sec:icphases} and study the constraints of a finite cylinder geometry. 
Our conclusions are summarized in Sec.~\ref{sec:conclusions}. 

\section{Model}
\label{sec:model}
We study the quantum-mechanical spin-1/2 Kitaev-$\Gamma$ model on the honeycomb lattice, which is described by the microscopic Hamiltonian 
\begin{equation}
\label{eq:model:hamiltonian}
H=\sum\limits_{\langle i, j \rangle_\gamma} K S_i^\gamma S_j^\gamma + \Gamma \left( S_i^\alpha S_j^\beta +  S_i^\beta S_j^\alpha \right) \,,
\end{equation}
where the sum runs over pairs of nearest neighbor lattice sites $i$ and $j$ which are connected by a lattice bond of type $\gamma\in x,y,z$ (with $\alpha,\beta$ denoting the remaining two orthogonal components). 
The exchange constant $K$ quantifies the diagonal, bond-directional couplings of Kitaev type, whereas $\Gamma$ captures symmetric off-diagonal interactions.
We parametrize the ratio of Kitaev and $\Gamma$ interactions by an angle $\alpha \in [0,2\pi)$ which is connected to the exchange constants via 
\begin{equation}
K=-\cos(\alpha) \quad\mathrm{and}\quad \Gamma=\sin(\alpha) \,.
\end{equation}
Throughout the remainder of this paper, all energy scales are reported in units of $\sqrt{K^2+\Gamma^2}$.

In terms of the parametrization described above, the regime $0<\alpha<\pi/2$, i.e. ferromagnetic (FM) Kitaev interactions $K<0$ in conjunction with antiferromagnetic (AFM) $\Gamma$ interactions $\Gamma>0$, is believed to be most relevant for the Kitaev material $\alpha$-\ce{RuCl3}~\cite{Winter2016}. 
Earlier zero-temperature model calculations have established, however, that this combination of exchange constants is not only particularly relevant for experiments, but it is also a challenging and controversial part of the phase diagram; various studies employing different methods have predicted different results. 
(i)~Variational Monte Carlo simulations suggest the existence of a sequence of small windows of ferromagnetic order, proximate Kitaev spin liquid, and incommensurate spiral order in vicinity to the Kitaev spin liquid around $\alpha=0$~\cite{Wang2019,Wang2020}. Zigzag order eventually manifests when the $\Gamma$ interaction becomes comparable in magnitude to the Kitaev contribution. 
(ii)~Tensor network representations of the model predict a slim region of ferromagnetic order and a nematic paramagnet near the Kitaev spin liquid, as well as more intricate magnetic order with a 6-site unit cell~\cite{Lee2020}. 
(iii)~Exact diagonalization studies on 24-site clusters point towards an extended spin liquid regime spanning all the way from the pure Kitaev spin liquid to pure $\Gamma$ interactions~\cite{Rau2014,Catuneanu2018}. 
(iv)~Recent density matrix renormalization group calculations find an extended nematic paramagnet regime in proximity to the Kitaev spin liquid~\cite{Gohlke2020}. 

Extensions of the pure Kitaev-$\Gamma$ model by additional detuning parameters have revealed additional insight into the structure of the phase diagram and the stability of competing phases; many of the aforementioned candidate ground states appear to have a phase boundary close the the pure Kitaev-$\Gamma$ model in this extended parameter space. 
Detuning parameters which have been studied in the past include competing Heisenberg interactions~\cite{Wang2019}, a spatial anisotropy in the interaction constants~\cite{Wang2020,Catuneanu2018}, a type of symmetric, off-diagonal exchange known as $\Gamma'$-interactions~\cite{Lee2020,Gohlke2020}, or simply a magnetic field~\cite{Lee2020,Chern2020a,Chern2020,Gohlke2020}.
Further related efforts to grow our understanding of the model also include studies of the dimensionally reduced one-dimensional Kitaev-$\Gamma$ chain~\cite{Yang2020,Yang2020a,Luo2020} or the Kitaev-$\Gamma$ ladder~\cite{Sorensen2021}. 
The close competition of many potential ground states has proven to render the numerical exploration of the phase space a formidable challenge.


\section{Functional renormalization group}
\label{sec:frg}

Our aim is to analyze the phase diagram of the model Hamiltonian Eq.~\eqref{eq:model:hamiltonian} within a pseudofermion functional renormalization group (pf-FRG) approach~\cite{Reuther2010,Buessen2019}. 
The pf-FRG approach is particularly intriguing in this situation, because it operates on an infinite representation of the honeycomb lattice, making it naturally compatible with incommensurate magnetic correlations, which have been suggested as one candidate among the potential ground states.
Moreover, the method does not rely on an explicit bias that would favor certain types of ground states, as would be the case e.g. in a mean-field construction or in the choice of initial states for variational approaches. 
In this section, we briefly outline our implementation of the pf-FRG approach, deferring the discussion of the resulting phase diagram to Sec.~\ref{sec:phasediagram}. 

We adopt an established formulation of the pf-FRG which is readily applicable to general time-reversal invariant Hamiltonians with two-spin interactions, including our model Hamiltonian Eq.~\eqref{eq:model:hamiltonian}~\cite{Buessen2019,Buessen2019a}. 
The approach is based on a parton construction in terms of spinful pseudofermions, $S_i^\mu=f^\dagger_{i,\alpha} \sigma^\mu_{\alpha\beta} f^{\phantom\dagger}_{i,\beta}$, with the indices $\alpha$ and $\beta$ implicitly summed over the two spin-1/2 basis states at every lattice site $i$, and a subsequent solution of the functional renormalization group flow equations for the resulting quartic fermion model within the well-developed framework of the fermionic functional renormalization group~\cite{Metzner2012,Wetterich1993}. 
The pseudofermion parton construction provides a faithful representation of the spin Hamiltonian if the half-filling constraint $\sum_\alpha f^\dagger_{i,\alpha} f^{\phantom\dagger}_{i,\alpha}=1$ is fulfilled at every lattice site $i$. 
Whereas the constraint can in principle be enforced by implementing an imaginary chemical potential $\mu_\mathrm{chem}=-\frac{i \pi T}{2}$~\cite{Popov1988}, such an approach is unfeasible numerically due to the concomitant reduction of symmetries of the Hamiltonian. 
Yet, it has been demonstrated that within the pf-FRG approach the constraint tends to be automatically fulfilled even for subtle incommensurate magnetic order~\cite{Baez2017,Kiese2020a,Buessen2018}. 
The implicit fulfillment of the constraint can be attributed to the fact that the pf-FRG approach is formulated at zero temperature and violations of the half-filling constraint, which lead to effective spin-0 defects, are thus energetically suppressed. 

The renormalization group (RG) flow equations are obtained by introducing a cutoff $\Lambda$ in the (Matsubara) frequency dependence of the bare propagator $G_0(\omega) \to G_0^\Lambda(\omega)=\frac{\theta\left( | \omega | - \Lambda \right)}{i \omega}$ (note the absence of a kinetic term in the bare propagator for pseudofermions); they form a set of differential equations which describes the evolution of higher-order fermionic interaction vertices under infinitesimal variation of the RG cutoff in the bare propagator. 
Per construction of the RG cutoff, the model remains unchanged at $\Lambda=0$, whereas in the opposite limit $\Lambda\to\infty$ the bare propagator vanishes, rendering the model particularly simple. 
In fact, when the bare propagator vanishes, the fermionic interaction vertices exactly match the frequency-independent interactions as defined in the model Hamiltonian Eq.~\eqref{eq:model:hamiltonian}, thus marking a well-defined starting point for the RG analysis. 

\begin{figure}
\includegraphics[width=\linewidth]{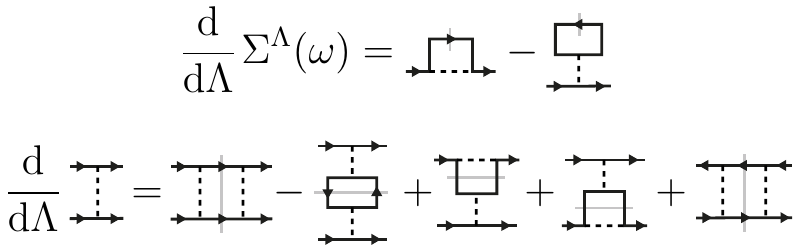}
\caption{{\bf Flow equations} for the cutoff-dependent self-energy $\Sigma^\Lambda(\omega)$ and the fermionic two-particle interaction vertices (dashed lines) $\Gamma_{i_1i_2}^\Lambda(1',2';1,2)$ in diagrammatic representation, where composite indices $k=(\omega_k,\alpha_k)$ comprise frequency and spin index. Lattice sites are preserved along solid lines. 
The self-energy is independent of lattice site and spin index. 
Slashed propagator lines resemble the single-scale propagator $S(\omega)=\frac{\delta\left( | \omega | - \Lambda \right)}{i \omega - \Sigma^\Lambda(\omega)}$, pairs of slashed propagator lines denote $G(\omega_1)S_\mathrm{kat}(\omega_2)+G(\omega_2)S_\mathrm{kat}(\omega_1)$ with $G(\omega)= \frac{\theta\left( | \omega | - \Lambda \right)}{i \omega - \Sigma^\Lambda(\omega)}$ and $S_\mathrm{kat}(\omega)=-\frac{\mathrm{d}}{\mathrm{d}\Lambda}G(\omega)$.
Details are provided in Ref.~\onlinecite{Buessen2019}.}
\label{fig:frg:flowequations}
\end{figure}

Imposing the cutoff, an exact hierarchy of coupled differential equations (flow equations) for fermionic $n$-particle vertices is generated at any order of $n$, which connects the known initial conditions at $\Lambda\to\infty$ to the physically relevant zero-cutoff limit. 
In general, this hierarchy of flow equations does not close, and it contains terms to arbitrary order of $n$.  
Within the pf-FRG approach, the hierarchy is then truncated such that it fully resolves the lattice site and frequency dependence of the self-energy and of the pseudofermionic two-particle interaction vertex, partially retaining feedback from the three-particle vertex~\cite{Reuther2010,Katanin2004}. 
The truncated flow equations are detailed in Ref.~\onlinecite{Buessen2019} and schematically shown in Fig.~\ref{fig:frg:flowequations}.
\begin{figure*}
\includegraphics[width=\linewidth]{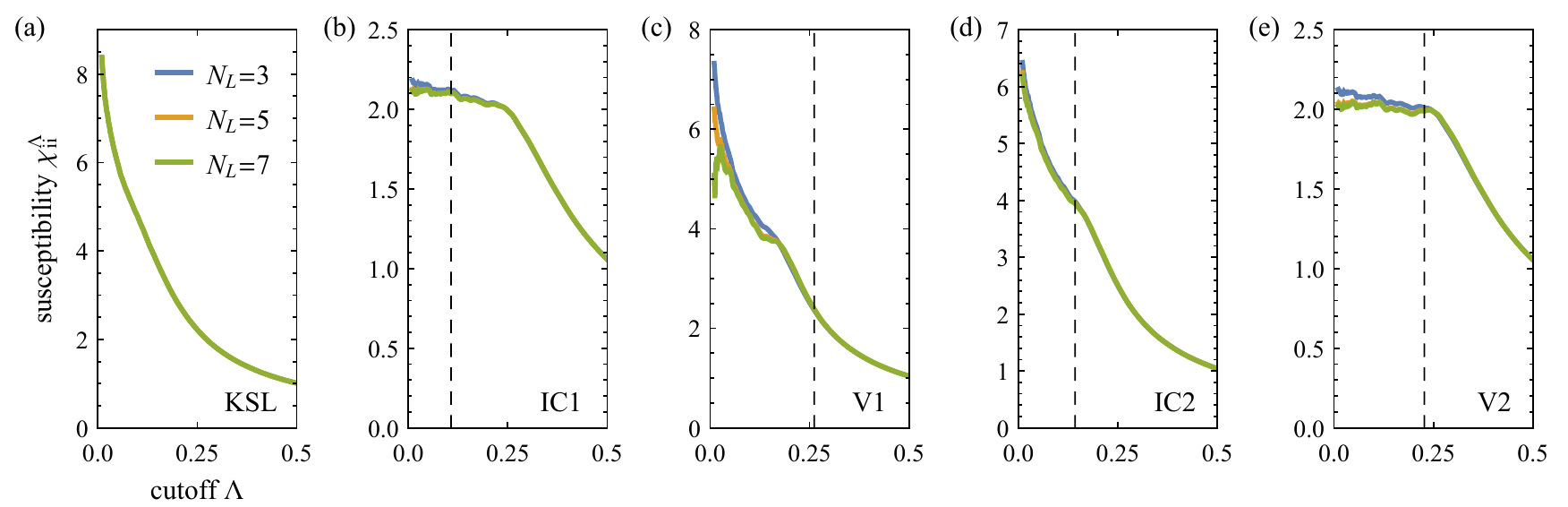}
\caption{{\bf Renormalization group flow} of the onsite magnetic susceptibility $\chi^\Lambda_{ii}$ for different lattice truncation ranges $N_L$. The dashed lines (if present) indicate the characteristic RG scale $\widetilde{\Lambda}_c$ which is associated with the occurrence of spontaneous symmetry breaking. The data displayed is in the (a)~Kitaev spin liquid (KSL) phase at $\alpha/\pi=0$, (b)~incommensurate phase 1 (IC1) at $\alpha/\pi=0.37$, (c)~vortex phase 1 (V1) at $\alpha/\pi=0.72$, (d)~incommensurate phase 2 (IC2) at $\alpha/\pi=1.30$, and (e)~vortex phase 2 (V2) at $\alpha/\pi=1.68$. }
\label{fig:phasediagram:characteristicscale_composite}
\end{figure*}

On this level of truncation, the pf-FRG formalism is understood to combine aspects of a large-S expansion~\cite{Baez2017} and an SU($N$) large-$N$ expansion~\cite{Buessen2018a,Roscher2018} on equal footing, recovering the mean-field exact results in the two separate limits. 

We formally consider the flow equations at zero temperature, where Matsubara frequencies become continuous. 
However, for the numerical solution of the flow equations we re-discretize the Matsubara frequency dependence and interpolate linearly in between the discrete grid points. 
In our implementation, the Matsubara frequency space is approximated by a mesh of $N_\omega=144$ points which are spaced logarithmically around the origin in the range $-250<\omega<250$. 

The dependence of the interaction vertices on the lattice site indices is treated as follows. 
We consider an infinite representation of the honeycomb lattice, but set any two-particle interactions to zero if the lattice sites involved are further apart than $N_L=7$ lattice bonds. 
This truncation scheme can be viewed as a series expansion in system size which eventually converges to a well-defined value when $N_L$ is chosen large enough. 
Most importantly, the truncation does not impose an artificial boundary onto the system. 
It has been shown that this approach is well suited to resolve incommensurate magnetic correlations~\cite{Buessen2018,Niggemann2020} -- which are typically difficult to simulate on finite lattice systems. 

Employing the aforementioned discretization schemes, one obtains a finite set of approximately $5 \times 10^7$ coupled integro-differential equations which describe the RG flow. 
The flow is evolved numerically from $\Lambda_\mathrm{UV}=500$, which exceeds any intrinsic energy scale of the system, to the effective low-energy theory. 
We stop the RG flow at $\Lambda_\mathrm{IR}=0.01$, below which no further qualitative changes are expected to occur. 

Throughout the evolution of the RG flow, instabilities may occur which signal spontaneous symmetry breaking and the onset of magnetic order~\cite{Reuther2010}. 
Such instabilities can be observed as nonanalyticities (kinks) in the flow of the magnetic susceptibility $\chi^\Lambda(\mathbf{k}) = \sum_{i,j} e^{-i \mathbf{k}\cdot (\mathbf{r}_i-\mathbf{r}_j)} \chi^\Lambda_{ij}$, 
where $\chi^\Lambda_{ij}$ are the real-space elastic spin correlations $\left< \mathbf{S}_i(\omega=0) \cdot \mathbf{S}_j(\omega=0) \right>$ which can be computed from the fermionic interaction vertices at any given RG scale~\cite{Reuther2010}. 
If no nonanalyticity is present down to the lowest energy scale $\Lambda_\mathrm{IR}$, the ground state of the system is assumed to retain its full symmetry -- which we loosely associate with spin liquid behavior. 

Locating the critical breakdown scale $\Lambda_c$ merely by visual inspection of the susceptibility curves introduces some uncertainty to the results. 
Attempts to identify more robust signatures of phase transitions within pf-FRG are therefore subject of current research~\cite{Keles2018,Kiese2020a,Kiese2020,Thoenniss2020}. 
Here, we employ an approach introduced in Ref.~\cite{Kiese2020a} which defines a characteristic RG scale $\widetilde{\Lambda}_c$ via the finite-size convergence of the system as follows.  
If one computes the on-site susceptibility $\chi^\Lambda_{ii}$ for different lattice truncation ranges $N_L=3,5,7$, one finds that the results converge beyond a truncation range which is comparable to the length scale of relevant interactions. 
In trivial paramagnets and spin liquid phases, only the short-range correlations are relevant and the susceptibility is typically converged already for a small truncation range $N_L=3$. 
Close to a magnetic ordering transition, however, long-range correlations proliferate and a discrepancy between results for $N_L=3$ and results for greater truncation ranges $N_L>3$ develops. 
The occurrence of such a sudden change in convergence behavior defines the characteristic RG scale $\widetilde{\Lambda}_c$ which, unlike the naive definition of $\Lambda_c$, can be computed systematically. 
On our level of numerical uncertainty, a minimum relative deviation of approximately 1\% between the susceptibility values for the different truncation ranges is found to be a good measure to determine $\widetilde{\Lambda}_c$.

Ultimately, our goal is to compute the $\omega=0$ component of the dynamic structure factor to characterize the phase diagram.
If the system does not undergo spontaneous symmetry breaking, we calculate the structure factor for the effective low-energy theory at the smallest cutoff scale $\Lambda_\mathrm{IR}$. 
If, on the other hand, a breakdown of the flow is observed, we calculate the structure factor at the characteristic scale $\widetilde{\Lambda}_c$. 
In the latter case, intensity maxima in the structure factor emerge already slightly above the transition scale and we can read off the nature of the incipient magnetic order. 
In fact, with the specific choice of the RG cutoff function in the pf-FRG approach, the RG scale $\Lambda$ can be interpreted as a temperature via a linear rescaling $\Lambda=\frac{2}{\pi}T$~\cite{Iqbal2016,Buessen2018a}, with the critical scale $\Lambda_c$ corresponding to the critical temperature of an ordering transition. 
As such, at finite RG cutoff, the pf-FRG approach resolves the combined effects of thermal fluctuations and quantum fluctuations~\cite{Buessen2018,Ghosh2019,Kiese2020}. 
We point out, however, that the RG flow is only meaningful until the first breakdown occurs, and we can thus only detect the first finite-temperature transition. 
If the model implies a multi-step ordering process, the lower-temperature phases cannot be resolved.


\section{Phase diagram}
\label{sec:phasediagram}
We now turn to the discussion of the phase diagram associated with our model Hamiltonian Eq.~\eqref{eq:model:hamiltonian} as a function of the angle $\alpha$, which parametrizes the ratio of Kitaev coupling and $\Gamma$ interactions. 
The main phase diagram, obtained within the pf-FRG approach, is summarized in Fig.~\ref{fig:introduction:phasediagram}. 

\begin{figure}
\includegraphics[width=\linewidth]{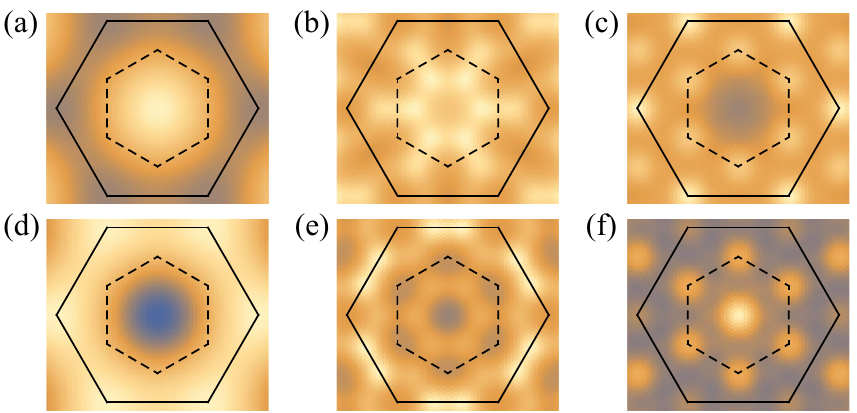}
\caption{{\bf Structure factors} of the Kitaev-$\Gamma$ model on the honeycomb lattice in the different phases. The dashed black lines indicate the first Brillouin zone, the extended Brillouin zone is denoted by the the solid lines. (a)~FM KSL phase, plotted at $\alpha \!=\! 0$. (b)~IC1 phase at $\alpha/\pi \!=\! 0.37$. (c)~V1 phase at $\alpha/\pi \!=\! 0.72$. (d)~AFM KSL phase at $\alpha/\pi \!=\! 1$. (e)~IC2 phase at $\alpha/\pi \!=\! 1.30$. (f)~V2 phase at $\alpha/\pi \!=\! 1.68$. The color code is normalized within each plot. All structure factors are plotted at the characteristic RG scale $\widetilde{\Lambda}_c$. }
\label{fig:phasediagram:sf_composite}
\end{figure}

In the vicinity of the FM Kitaev point ($\alpha/\pi=0$), we naturally observe the manifestation of a paramagnetic ground state -- the ferromagnetic Kitaev spin liquid (KSL). 
The concomitant absence of spontaneous symmetry breaking is indicated by a smooth RG flow of the magnetic susceptibility down to the lowest cutoff $\Lambda_\mathrm{IR}$, as depicted in Fig.~\ref{fig:phasediagram:characteristicscale_composite}a. 
The corresponding momentum-resolved structure factor $\chi(\mathbf{k})$ of the FM KSL is displayed in Fig.~\ref{fig:phasediagram:sf_composite}a, revealing a broad maximum around the Brillouin zone (BZ) center. 
From the analytical solution in the pure Kitaev limit, it is known that this broad maximum results from the superposition of three cosine profiles associated with nearest-neighbor correlations along the $x$, $y$, and $z$-type bonds. 

The KSL phase remains stable upon inclusion of small AFM $\Gamma$ interactions. 
When the strength of AFM $\Gamma$ interactions is increased further, the spin liquid eventually becomes unstable and a finite characteristic scale $\widetilde{\Lambda}_c$ is observed beyond $\alpha/\pi > 0.05$. 
For a brief window, $0.05<\alpha/\pi<0.15$, the dominant wave vector remains indifferent from our findings for the FM KSL phase -- peaking at the BZ center -- which, in combination with the finite characteristic scale $\widetilde{\Lambda}_c$, suggests the formation of ferromagnetic (FM) order. 
The manifestation of a small region of ferromagnetic order has previously also been observed in variational Monte Carlo calculations~\cite{Wang2019} and tensor network optimizations~\cite{Lee2020}. 
In the latter study it was found that the FM order is in close energetic proximity to competing configurations, being selected as the ground state only by an energy difference $\Delta E \sim \mathcal{O}(10^{-4})$. 
Such near-degeneracy of potential ground state configurations in this region of the phase space could contribute to our finding that the finite characteristic scale $\widetilde{\Lambda}_c$ appears to connect smoothly to the neighboring phase upon further increasing the angle $\alpha/\pi$. 

In the range $0.15 < \alpha/\pi < 0.61$, we observe a broad regime in which the model exhibits incommensurate magnetic correlations. 
We refer to this regime as the first incommensurate phase~(IC1). 
While the characteristic scale $\widetilde{\Lambda}_c$ is non-zero, see Fig.~\ref{fig:phasediagram:characteristicscale_composite}b, it remains relatively small compared to the characteristic scale of other magnetically ordered phases in the phase diagram (c.f. Fig.~\ref{fig:introduction:phasediagram}), suggesting that the overall ordering tendency in this regime is only weak. 
The characteristic wave vector which describes the evolution of incommensurate spiral order lies on the line connecting the BZ center and the edges of the first BZ ($M$-points), see Fig.~\ref{fig:phasediagram:sf_composite}b. 
Its precise position along the line shifts continuously within the phase; we discuss the IC1 phase in greater detail in Sec.~\ref{sec:icphases}. 

Around $0.61 < \alpha/\pi < 0.94$, where the Kitaev interaction and $\Gamma$ interaction are both finite and AFM, the system transitions into conventional magnetic order, signaled by a distinct breakdown of the RG flow as displayed in Fig.~\ref{fig:phasediagram:characteristicscale_composite}c. 
The associated structure factor shows dominant peaks on the corners of the extended BZ and sub-dominant peaks in close vicinity to the corners of the first BZ ($K$-points), see Fig.~\ref{fig:phasediagram:sf_composite}c. 
We refer to this phase as the first vortex (V1) phase; it nucleates around the special point $\alpha_\mathrm{FM}/\pi=0.75$ which is dual to the Heisenberg ferromagnet via an underlying six-sublattice duality transformation~\cite{Chaloupka2015}. 
Its existence has previously been established also in ED calculations~\cite{Rau2014} and DMRG calculations~\cite{Gohlke2018}. 
However, in our calculations we also observe a small drift of the sub-dominant peak position, $\mathbf{k}_s$, near the $K$-points.
Upon entry at the lower phase boundary, the peak is slightly dislocated from the $K$-point towards the BZ center and it subsequently shifts outwards as $\alpha$ is increased; it crosses the $K$-point approximately when $\alpha$ is tuned to the center of the V1 phase, near the special point $\alpha_\mathrm{FM}$. 
The overall shift is observed to be small, on the order $|\delta \mathbf{k}_s|/|\mathbf{k}_s|\approx 0.05$. 

In the vicinity of the AFM Kitaev point, magnetic order is eventually destabilized in favor of the AFM Kitaev spin liquid, which forms the ground state between $0.94 < \alpha/\pi < 1.09$. 
In analogy to the FM KSL phase, the corresponding structure factor reveals broad maxima around the corners of the extended BZ, resulting from the superposition of three cosine profiles (Fig.~\ref{fig:phasediagram:sf_composite}d). 

The sequence of phases which follow in the second half of the phase diagram, i.e. for $\alpha/\pi >1$, is remarkably similar to the first half. 
A small region of antiferromagnetic order ($1.09<\alpha/\pi <1.12$) separates the AFM KSL from a second phase of incommensurate spiral order, the IC2 phase ($1.12<\alpha/\pi <1.34$). 
In the latter phase, the dominant wave vectors characterizing the incommensurate order continuously shift on the edges of the extended Brillouin zone, between the $\Gamma'$-points and the $M'$-points, see Fig.~\ref{fig:phasediagram:sf_composite}e and Sec.~\ref{sec:icphases} for a more detailed discussion. 
Subsequently, for a broad parameter regime with Kitaev and $\Gamma$ interactions being mostly ferromagnetic, a second vortex phase is stabilized (V2, $1.38<\alpha/\pi <1.98$), with the possibility of a slim paramagnetic regime in between the IC2 and V2 phases. 
The V2 phase is associated with a structure factor that is peaked at the Brillouin zone center, with sub-leading maxima appearing on the corners of the first BZ as well as of the extended BZ (Fig.~\ref{fig:phasediagram:sf_composite}f). 
It nucleates around the special point $\alpha_\mathrm{AFM}/\pi=1.75$, which is dual to the Heisenberg antiferromagnet via a six-sublattice transformation~\cite{Chaloupka2015}. 
Similar to our observations in the V1 phase, the sub-leading peaks near the $K$-point in the V2 phase show a small drift, whereas the peaks at the $\Gamma$-point and the $\Gamma'$-points remain immobile. 

Our overall phase diagram holds remarkable similarity to previous zero-temperature ED calculations, which have been performed across the full parameter space on 24-site clusters~\cite{Rau2014}. 
We emphasize two important benchmarks: 
(i) First, we compare the stability of the Kitaev spin liquid under perturbation of small $\Gamma$ interactions. 
Since any finite spin correlations in the unperturbed Kitaev spin liquid are restricted to nearest neighbors only, one can expect that finite-size effects in ED calculations remain small in the vicinity of the pure Kitaev points, and ED results may thus provide good numerical guidance. 
We find that our pf-FRG calculations correctly reproduce the stability of the Kitaev spin liquid under perturbations of AFM $\Gamma$ interactions, which is approximately the same for the FM KSL and the AFM KSL~\cite{Rau2014,Gohlke2018}, whereas under perturbation of FM $\Gamma$ interactions the AFM KSL is significantly more robust than the FM KSL~\cite{Rau2014}. 
(ii) Second, we focus on the formation of the magnetic vortex phases V1 and V2. Since these phases nucleate around the special points $\alpha_\mathrm{FM}$ and $\alpha_\mathrm{AFM}$, respectively, where it is known that the magnetic order has a six-site unit cell~\cite{Chaloupka2015}, they can be well captured in ED calculations~\cite{Rau2014} and in DMRG calculations~\cite{Gohlke2018}. 
Similar to those ED and DMRG results, our pf-FRG calculations correctly capture the extended domain of stability for the V2 phase and the smaller parametric extent of the V1 phase.

Our identification of the incommensurate phases IC1 and IC2, however, deviates from previous ED and DMRG results. 
Furthermore, we observed a small, sub-dominant incommensurability effect in the vortex phases. 
While the incommensurate phases have not been reported in DMRG calculations, the formation of spiral order has been reported in ED calculations in parametric regimes where pf-FRG observes incommensurate order -- albeit in ED calculations the spiral pitch vectors are naturally locked to momentum points which are compatible with the finite system size~\cite{Rau2014}. 
In this comparison, it is important to keep in mind that pf-FRG calculations only resolve the system at RG scales down to the flow breakdown. 
A potential multi-step ordering process below the critical scale $\Lambda_c$ cannot be resolved. 
In particular, the results at a finite RG scale $\Lambda>0$ incorporate the joint effect of thermal and quantum fluctuations and related order-by-disorder effects, which can be decisively different from results obtained in zero-temperature ED or DMRG calculations.


\section{Incommensurate phases}
\label{sec:icphases}
\begin{figure}
\includegraphics[width=\linewidth]{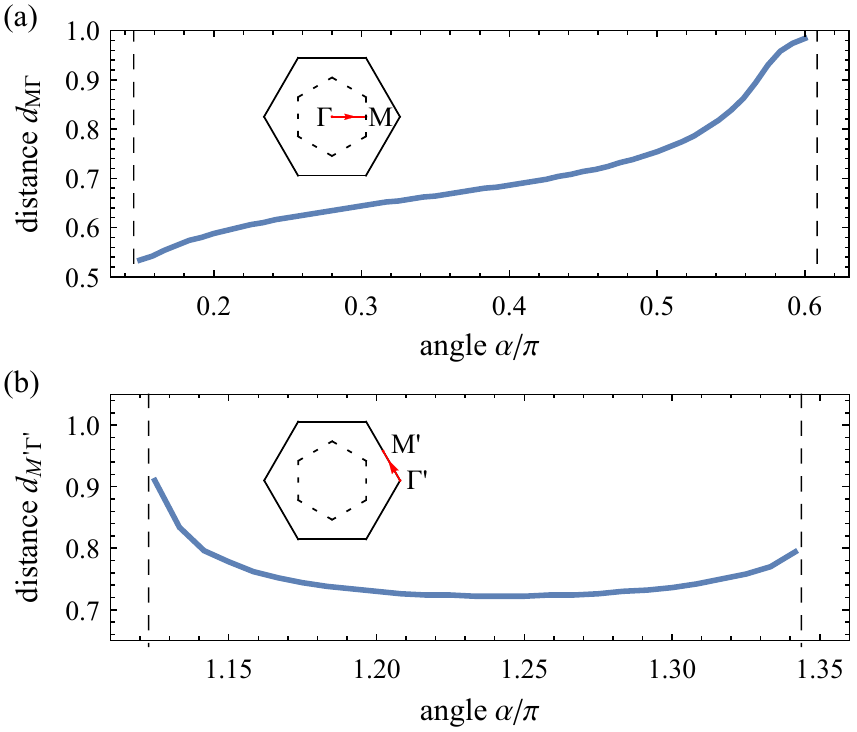}
\caption{{\bf Incommensurate ordering vectors} in the two incommensurate phases IC1 and IC2. The plots show the position $\mathbf{k}_\mathrm{max}$ of the structure factor peak within the Brillouin zone. 
(a) The peak in the IC1 phase moves continuously on the momentum-space line (indicated as a red arrow in the inset) from the BZ center ($\Gamma$-point, corresponds to relative distance $d_{M\Gamma}=0$) to the BZ edge ($M$-point, corresponds to relative distance $d_{M\Gamma}=1$) as a function of the angle $\alpha$. The dashed lines indicate the phase boundaries of the IC1 phase. 
(b) Relative distance of the peak in the IC2 phase between the $\Gamma'$-point and the $M'$-point.}
\label{fig:icphases:kEvolution_composite}
\end{figure}
\begin{figure}
\includegraphics[width=\linewidth]{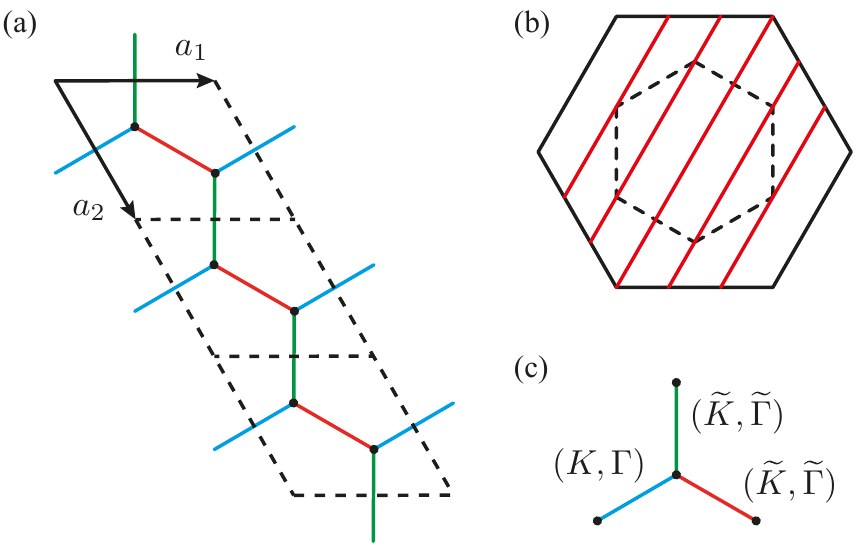}
\caption{{\bf Finite cylinder geometry}. (a) The rhombic unit cell is indicated by dashed lines. The cylinder extends infinitely in the $a_1$-direction and is assumed to have periodic boundary conditions in the $a_2$-direction. (b) Red lines indicate lines of allowed momenta within the extended BZ for a circumference of 6 lattice sites (3 unit cells). Dashed lines indicate the first BZ. (c) Explicit breaking of symmetry by modulating the exchange constants $(\widetilde{K},\widetilde{\Gamma}) \neq (K,\Gamma)$ along the $x$ and $y$-bonds. }
\label{fig:icphases:cylindergeometry}
\end{figure}
We now turn our attention to the two incommensurate phases, IC1 and IC2. 
Conventional magnetic long-range order is usually characterized by a single peak (and symmetry-related copies) in the spin structure factor which defines the ordering vector in momentum space, cf. the structure factors of the V1 and V2 vortex phases displayed in Fig.~\ref{fig:phasediagram:sf_composite}. 
Typically, this ordering vector remains constant within the phase. 
This is, however, qualitatively different in the incommensurate spiral phases IC1 and IC2: Here, the associated ordering vector evolves continuously within the phase upon variation of the underlying exchange constants, yielding an ordering vector~$\mathbf{k}_\mathrm{max}$ that is dependent on the angle $\alpha$. 

In the IC1 phase, the ordering vector shifts within the first BZ along the line which connects the BZ center (the $\Gamma$-point) to the mid-point of the BZ edge (the $M$-point). 
Upon entry of the IC1 phase at $\alpha/\pi \approx 0.15$, the ordering vector $\mathbf{k}_\mathrm{max}$ is located approximately in the middle between the $\Gamma$-point and the $M$-point, corresponding to a relative distance $d_{M\Gamma} \approx 0.5$, where $d_{M\Gamma} = \left| \mathbf{k}_\mathrm{max}-\mathbf{k}_\Gamma \right| / \left| \mathbf{k}_M-\mathbf{k}_\Gamma \right|$ with $\mathbf{k}_\Gamma$ the position of the $\Gamma$-point, $\mathbf{k}_M$ the position of the $M$-point, and $|.|$ measures the usual Euclidean distance. 
When $\alpha$ is increased, the structure factor peak monotonically shifts outwards until it approaches the $M$-point ($d_{M\Gamma}=1$) near the phase boundary at $\alpha/\pi \approx 0.61$. 
The precise trajectory of the peak is shown in Fig.~\ref{fig:icphases:kEvolution_composite}a. 

Similarly, in the IC2 phase, the ordering vector continuously shifts along the edge of the extended BZ. 
Unlike in the IC1 phase, however, the shift is non-monotonic. 
As shown in Fig.~\ref{fig:icphases:kEvolution_composite}b, at the lower phase boundary ($\alpha / \pi \approx 1.12$) the peak starts out near the mid-point of the extended BZ edge ($M'$-point), $d_{M'\Gamma'} \approx 0.9$, retracts closer towards the extended BZ corner ($\Gamma'$-point) deep inside the IC2 phase, $d_{M'\Gamma'} \approx 0.7$, before extending again towards the $M'$-point near the upper phase boundary ($\alpha / \pi \approx 1.34$), $d_{M'\Gamma'} \approx 0.8$.

A continuous shifting of the incommensurate magnetic ordering vector can often be cumbersome for numerical studies on finite lattices: If the lattice is treated with open boundary conditions, boundary effects can become sizable. 
If, on the other hand, periodic boundary conditions are applied, only a restricted set of momentum points within the Brillouin zone can be resolved (c.f. Fig.~\ref{fig:icphases:cylindergeometry}b). 
Within the pf-FRG approach, however, we were able to treat the system in the thermodynamic limit where the continuous momentum dependence of the ordering vector is accessible. 
We now complement our calculations on the infinite system with additional pf-FRG calculations on a semi-infinite cylinder in an attempt to estimate the impact of periodic boundary conditions. 
We chose a rhombic unit cell as depicted in Fig.~\ref{fig:icphases:cylindergeometry}a and consider a cylinder with a 6-site (3 unit cell) circumference, which has also been employed in previous DMRG calculations~\cite{Gohlke2018,Gohlke2020}. 
Potential anisotropies in the ground state which may result from the geometric constraints can be quantified via the deviation of the nearest-neighbor spin correlation $\chi^\Lambda_{ij}$ on a $\mu$-bond ($\mu=x,y,z$) from the mean value across all three types of bonds. 
We thus define the anisotropy measure 
\begin{equation}
\zeta^\Lambda_\mu = \frac{3 \chi^\Lambda_\mu}{\chi^\Lambda_x + \chi^\Lambda_y + \chi^\Lambda_z} - 1 \,,
\end{equation}
where $\chi^\Lambda_\mu$ is a shorthand notation for the spin correlation $\chi^\Lambda_{ij}$ with $i$ and $j$ being nearest neighbors along a $\mu$-bond. 

\begin{figure}
\includegraphics[width=\linewidth]{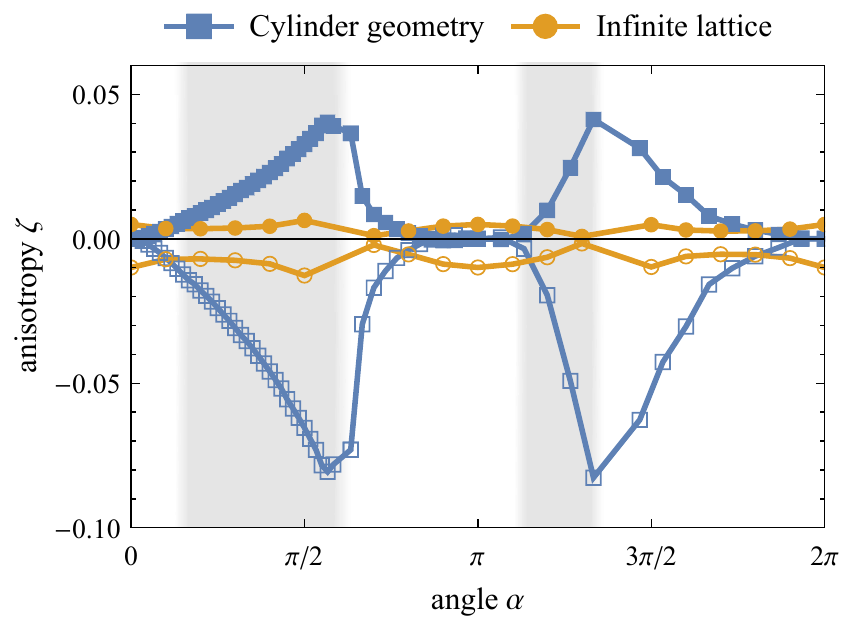}
\caption{{\bf Bond anisotropy} on a cylinder with rhombic unit cell and 6-site circumference as a function of the angle $\alpha$ (blue curve), plotted at the characteristic scale $\widetilde{\Lambda}_c$. Solid markers denote $\zeta_x = \zeta_y$, open markers denote $\zeta_z$. The grey shaded regions correspond to the incommensurate phases IC1 and IC2, respectively. 
The yellow curve shows the anisotropy on an infinite lattice with small explicit symmetry breaking in the exchange constants, $\epsilon = 0.01$ (see text for details).}
\label{fig:icphases:anisotropy}
\end{figure}

Performing the pf-FRG calculations on the cylinder geometry, we find that the characteristic RG scale $\widetilde{\Lambda}_c$ remains qualitatively unchanged across the full phase diagram. 
We further observe that the anisotropy in the Kitaev spin liquid remains negligible, since the correlation length is short compared to the cylinder circumference. 
In the magnetically ordered phases, on the other hand, a finite anisotropy develops, see Fig.~\ref{fig:icphases:anisotropy}. 
Similar to the phase diagram in the thermodynamic limit, there exists a striking resemblance between the first half of the phase diagram $0<\alpha/\pi<1$, and the second half $1<\alpha/\pi<2$: Being negligible in the FM (AFM) KSL phase, the anisotropy becomes finite upon entering the incommensurate phase IC1 (IC2), continuously growing larger as the angle $\alpha$ is increased, reaching its maximum near the phase transition into the V1 (V2) phase. 
The finite extent of the anisotropy into the V1 and V2 ordered phases may seem surprising, since the phases nucleate around the special points $\alpha_\mathrm{FM}$ and $\alpha_\mathrm{AFM}$, respectively, where it is known that the underlying magnetic order is compatible with the rhombic 6-site cylinder geometry~\cite{Chaloupka2015}. 
Yet, our observation of finite anisotropy is in agreement with the weak incommensurate shift of sub-leading peaks in the corresponding structure factors away from the special points, which we discussed in Sec.~\ref{sec:phasediagram}. 
This incommensurate displacement is largest at the phase boundary and vanishes near the special points -- similarly, we observe that the anisotropy vanishes as the special points are approached, cf. Fig.~\ref{fig:icphases:anisotropy}. 

Having established the different proliferation of anisotropy across the various phases on a cylinder geometry, we now assess the absolute magnitude of the anisotropy. 
To this end, we compare our results on the cylinder geometry to a system in the thermodynamic limit, where we explicitly break the rotational symmetry in the Hamiltonian. 
In the model Hamiltonian Eq.~\eqref{eq:model:hamiltonian} we thus replace the exchange constants $K$ and $\Gamma$ on the $x$-bonds and $y$-bonds by modified interactions
\begin{equation}
\begin{pmatrix} \widetilde{K} \\ \widetilde{\Gamma} \end{pmatrix} = (1+\epsilon)\times \begin{pmatrix} K \\ \Gamma \end{pmatrix} \,,
\end{equation}
where $\epsilon$ parametrizes the strength of the initial symmetry breaking (cf. Fig.~\ref{fig:icphases:cylindergeometry}c). 
For a negligible symmetry breaking ($\epsilon = 0.01$) in the initial model we do not observe a substantial RG flow towards an enhanced anisotropy, as shown in Fig.~\ref{fig:icphases:anisotropy}. 
Instead, the anisotropy measure $\zeta^{\widetilde{\Lambda}_c}_\mu$, plotted at the characteristic scale $\widetilde{\Lambda}_c$, remains small and with little variation across the different phases when compared to the anisotropy arising on the cylinder geometry.
We find that the level of anisotropy induced by the cylinder geometry would require a more substantial initial symmetry-breaking of approximately $\epsilon \approx 0.05$ -- though we note that the absolute magnitude of the anisotropy measure is of limited meaning, since within the pf-FRG approach we can only trace its evolution up to the phase transition at the critical scale $\Lambda_c$, and we cannot estimate the value it assumes deep within the ordered phase. 
Regardless, these results suggest that the tendency of the ground state configuration to become anisotropic in the incommensurate phases IC1 and IC2, but also in the vortex phases V1 and V2, is significantly enhanced on the cylinder geometry, and a careful finite-size analysis is crucial in this setting.


\section{Conclusions}
\label{sec:conclusions}
We have mapped out the phase diagram of the spin-1/2 Kitaev-$\Gamma$ model defined in Eq.~\eqref{eq:model:hamiltonian} as a function of the angle $\alpha$, which parametrizes the ratio of Kitaev and $\Gamma$ interactions, utilizing a pf-FRG approach which is sensitive to both, quantum and thermal fluctuations. 
Away from the pure Kitaev limit, we have identified four extended, magnetically ordered phases which are stabilized by non-vanishing $\Gamma$ interactions at finite temperature: the incommensurate spiral phases IC1 and IC2, as well as the vortex phases V1 and V2; we further identified two narrow phases with predominantly FM and AFM magnetic correlations, respectively. 

In comparison to the pronounced breakdown of the RG flow in in the vortex phases V1 and V2, the breakdown in the incommensurate phases IC1 and IC2 is less distinct and happens at a low characteristic scale $\widetilde{\Lambda}_c$, which we associate with a low ordering temperature. 
In light of previous zero-temperature ED calculations, which revealed a close competition of many competing phases in the parameter regions in question~\cite{Rau2014}, the observation of a suppressed ordering temperature seems plausible. 
However, some uncertainty about the interpretation of the breakdown of the RG flow remains, since it does not manifest as a true divergence. 
The classification of a finite kink in the RG flow as either a genuine flow breakdown, or merely a non-analytic feature, introduces some ambiguity  -- which is an inherent limitation of the pf-FRG approach. 
In this work, we followed a classification scheme for the flow breakdown which is based on the finite-size scaling behavior of the RG flow and has proven useful in previous studies~\cite{Kiese2020a}. 
Yet, a more rigorous classification would be desirable, since even the finite-size scaling approach cannot fully resolve all ambiguity and could, in principle, overestimate magnetic ordering tendencies. 
Such a rigorous approach seems within reach: Only recently, a truncation of the flow equations has been proposed which goes beyond the standard pf-FRG truncation employed here. 
For Heisenberg-like quantum spin models, it was demonstrated that this so-called \emph{multiloop} pf-FRG scheme can significantly sharpen the signature of the flow breakdown and consequently reduce uncertainty~\cite{Kiese2020,Thoenniss2020}. 
It would thus be an exciting project for future research to generalize the multiloop truncation to microscopic spin models with reduced symmetry, akin to the one studied here. 

Regardless of their eventual stability, we were able to identify the leading magnetic ordering channels in the thermodynamic limit and explicitly resolve the role of incommensurate magnetic correlations, which could only be speculated on in previous finite-size calculations.
Most notably, we showed that large portions of the phase diagram, namely the IC1 and IC2 regions, are governed by incommensurate magnetic correlations which continuously evolve under variation of the angle $\alpha$. 
On a more subtle note, in the vortex phase V1 (V2), we observed a small incommensurate drift of sub-leading peaks around the corners of the Brillouin zone which vanishes near the special point $\alpha_\mathrm{FM}$ ($\alpha_\mathrm{AFM}$), where the Kitaev-$\Gamma$ model becomes dual to the Heisenberg ferromagnet (antiferromagnet). 

In order to aid the interpretation of our results and to connect to previous zero-temperature numerical results on finite lattices, obtained e.g. by ED or DMRG calculations, we further performed pf-FRG calculations on a semi-infinite cylinder. 
We found that a significant lattice anisotropy in the two-spin correlation function arises if the underlying magnetic order is incompatible with the periodic boundary conditions of the cylinder geometry. 
The latter is naturally the case for incommensurate magnetic order, i.e. in the incommensurate phases IC1 and IC2. 
Yet, we also observed the anisotropy to extend into the vortex phases V1 and V2, which we associate with the weak incommensurability effect of sub-dominant peaks showing a small drift near the corners of the Brillouin zone. The anisotropy vanishes concurrently with the displacement of sub-dominant peaks when the special point $\alpha_\mathrm{FM}$ or $\alpha_\mathrm{AFM}$, respectively, is approached. 

Going beyond the scope of our current work, where we focused on the minimal Kitaev-$\Gamma$ model for a general set of Kitaev honeycomb materials, it would be instructive to extend the analysis to cover more specific material parameters, informed by \emph{ab initio} calculations, and potentially including Heisenberg and $\Gamma'$ interactions. 
Incorporating these extra interactions in the pf-FRG approach is straight-forward. 
Eventually, it would also be interesting to study the related model on the hyperhoneycomb lattice, which is relevant to the Kitaev material $\beta$-\ce{Li2IrO3} -- in a three-dimensional setting, the pf-FRG approach would have a decisive advantage over other numerical techniques like ED or DMRG, which are severely limited by finite system sizes.


\begin{acknowledgments}
We thank Matthias Gohlke for helpful discussions. 
Parts of the numerical simulations were performed on the Cedar cluster, hosted by WestGrid and Compute Canada. 
This work is supported by the NSERC of Canada and the Center for Quantum Materials at the University of Toronto. 
YBK is also supported by the Killam Research Fellowship from the Canada Council for the Arts. 
\end{acknowledgments}


\bibliography{honeycombKG}

\end{document}